\documentclass[fleqn,usenatbib]{mnras}
\usepackage{newtxtext,newtxmath}
\usepackage[T1]{fontenc}
\usepackage{ae,aecompl}
\usepackage{graphicx}	% Including figure files
\usepackage{amsmath}	% Advanced maths commands
\usepackage{amssymb}	% Extra maths symbols
\title[Breathing oscillations in a thin disk]{Breathing Oscillations in a Global Simulation of a Thin Accretion Disk}

\author[Mishra et al.]{
Bhupendra Mishra,$^{1,4}$\thanks{E-mail: bhupendra.mishra@jila.colorado.edu}
W{\l}odek Klu\'zniak,$^{2}$
P. Chris Fragile$^{3,4}$
\\
$^{1}$JILA, University of Colorado and National Institute of Standards and Technology, 440 UCB, Boulder, CO 80309-0440, USA\\
$^{2}$Nicolaus Copernicus Astronomical Center, Bartycka 18, Warsaw, 00-716, Poland\\
$^{3}$Department of Physics and Astronomy, College of Charleston, Charleston, SC 29424, USA \\
$^{4}$Kavli Institute for Theoretical Physics, Santa Barbara, CA.
}
\date{Accepted XXX. Received YYY; in original form ZZZ}

\pubyear{2018}

\begin{document}
\label{firstpage}
\pagerange{\pageref{firstpage}--\pageref{lastpage}}
\maketitle

\begin{abstract}
We study the oscillations of an axisymmetric, viscous, radiative, general relativistic hydrodynamical simulation of a geometrically thin disk around a non-rotating, $6.62\,M_\odot$ black hole. The numerical setup is initialized with a Novikov-Thorne, gas-pressure-dominated accretion disk, with an initial mass accretion rate of $\dot{m} = 0.01\,L_\mathrm{Edd}/c^2$ (where $L_\mathrm{Edd}$ is the Eddington luminosity and $c$ is the speed of light). Viscosity is treated with the $\alpha$-prescription. The simulation was evolved for about $1000$ Keplerian orbital periods at three Schwarzschild radii (ISCO radius). Power density spectra of the radial and vertical fluid velocity components, the total (gas $+$ radiation) midplane pressure, and the vertical component of radiative flux from the photosphere, all reveal strong power at the local breathing oscillation frequency.  The first, second and third harmonics of the breathing oscillation are also clearly seen in the data. We quantify the properties of these oscillations  by extracting eigenfunctions of the radial and vertical velocity components and total pressure. This confirms that these oscillations are associated with breathing motion.
\end{abstract}

\begin{keywords}
accretion, accretion disks -- variability -- X-rays:binaries
\end{keywords}

%%%%%%%%%%%%%%%%%%%%%%%%%%%%%%%%%%%%%%%%%%%%%%%%%%

%%%%%%%%%%%%%%%%% BODY OF PAPER %%%%%%%%%%%%%%%%%%
\section{Introduction}
\label{intro}
%Background%%%%%%
Accretion disks of various geometries (thin, slim and thick) have been proposed over the past few decades \citep{Pringle1972,Shakura73,Abramowicz1988,Abramowicz1978}, and they are widely used to model a wide range of astrophysical systems. Particularly in the context of X-ray variability, salient properties of X-ray binaries, such as their outbursts and the spectra of the bright state, have been successfully modeled with accretion disks. Accreting relativistic sources (supermassive black holes as well as stellar mass black holes, and neutron stars) often show X-ray variability, including robust quasi-periodicities. In particular, numerous observations of quasi-periodic oscillations (QPOs) in the X-ray flux of X-ray binaries harboring neutron stars or black holes have been reported 
\citep[reviewed in][]{vanderklis2000, Remillard06}.
It is hoped that studying and modeling such characteristic variability will advance our understanding of accretion disks, much as studies of stellar pulsations inform the theory of stellar structure, and helioseismology provides a direct probe of the solar interior. 
Of particular interest are high frequency QPOs, which have a frequency range of $40-450 \mathrm{Hz}$ in black hole sources \citep{Morgan97,Remillard99,Remillard02, Belloni2001, Miller2001, Strohmayer01, Altamirano2012, Belloni2012}. As described below, while many analytic models of such QPOs have been proposed, it is challenging to find their counterparts in numerical simulations of accretion disks.

Many models, not necessarily related to oscillations of thin disks, have been invoked to explain the observed highest frequency X-ray variability (the ``twin-peak" kHz QPOs of neutron star low mass X-ray binaries and their hectoHertz counterparts in the black hole X-ray binaries)---e.g., the so called relativistic precession model \citep{Stella98,Stella99}, or the resonance model \citep{Kluzniak01,Abramowicz01}---as well as lower frequency variability such as the C-type QPOs  of black hole X-ray binaries \citep[e.g. the precessing torus of][]{Ingram2009}. Among those models, eigenmodes of accretion tori and geometrically thick disks have been studied extensively, both numerically and analytically
\citep{Rezzolla03,Zanotti03,Bursa04,Zanotti05,Blaes06,Fragile16,Mazur16,Mishra17}.
 Here, we will be focusing on thin accretion disks, for which there exists a well developed analytic theory of normal modes \citep[reviewed in][]{Wagoner99, Kato2001}, but the numerical simulations of which are especially challenging. Observationally, X-ray variability has not been seen when accretion is in the high soft or disk state. However, our numerical simulation suggests that there are intrinsic disk oscillations which do not appear in the computed spectra. This could be possible due to two reasons: first, the optical thickness of such disks may prevent the variability from reaching the disk photosphere; and second, these oscillations are local rather than global. 
 
 There are only a handful of global numerical simulations of thin accretion disks in the context of black holes \citep{ONeill09, Reynolds09, Mishra16, Sadowski16,Morales2017,Hogg2018}. 
The results described below are based on our radiative, viscous, relativistic hydrodynamic simulations of geometrically thin black hole accretion disks \citep{Fragile2018}. 
 In this paper we unambiguously identify the presence in the simulations of a particular type of local, thin-disk oscillations, corresponding to the well studied breathing modes of accretion tori \citep{Blaes06} and spherical (or plane-parallel) atmospheres \citep{deepika2017}.
 
It has been found that the breathing mode of oscillations gets easily excited in hydrodynamical numerical simulations of geometrically thick disks (tori) \citep{Mishra17}.  In \citet{Mishra17}, we performed a set of inviscid hydrodynamical thick disk simulations to study the high frequency QPOs. The equilibrium torus was perturbed, and the resulting motions were followed in the numerical run. All of the simulations in that set showed the breathing mode together with oscillations of various other modes \citep[radial, vertical, plus, and X-modes discussed in][]{Blaes06}. In those thick disk simulations, we found  the breathing mode to be excited irrespective of initial velocity perturbations (radial, vertical or combination of two). 

In thin disks, there is no global breathing mode, but locally the disk may execute analogous motions in which the thickness of the disk, as well as its pressure, undergo harmonic variations. However, such breathing oscillations were not seen in the geometrically thin disk simulations of \citet{ONeill09} and \citet{Reynolds09}. Unlike those authors, we include radiation (and approximate radiative transfer) in our simulations, and for the first time we report the presence of breathing oscillations in global thin disk simulations.  The frequency is in agreement with the predicted frequency for local breathing oscillations in the plane-parallel approximation \citet{silb2001, deepika2017}.  Thus, a fundamental theoretical prediction is confirmed by us in a fully non-linear, numerical simulation, which includes general relativity and radiation.

This paper is organized as follows, in Section~\ref{setup} the numerical setup is described. In Section~\ref{outcome}, we report the preliminary findings of this simulation. Section~\ref{results} describes diagnostics, power density spectra and eigenfunctions. Section~\ref{discuss} discusses the reported results. We conclude our study in Section~\ref{conclusions}. The metric signature in the paper is $-+++$. The numerical computations are carried out in Kerr-Schild coordinates with black hole mass $M= 6.62\,M_\odot$.
The simulations are performed using the general relativistic radiative (magneto) hydrodynamical code {\it Cosmos++} \citep{Anninos05, Fragile12, Fragile14}.

%%%%%%%%%%%%%%%%%%%%%%%%%%%%%%%%%%%%%%%%%%%%%%%%%%%%%%%%
%%%%%%%%%%%%%%%%%%%%%%%%%%%%%%%%%%%%%%%%%%%%%%%%%%%%%%%%

 \section{Numerical setup}
 \label{setup}
\begin{figure*}
\includegraphics[width=\columnwidth]{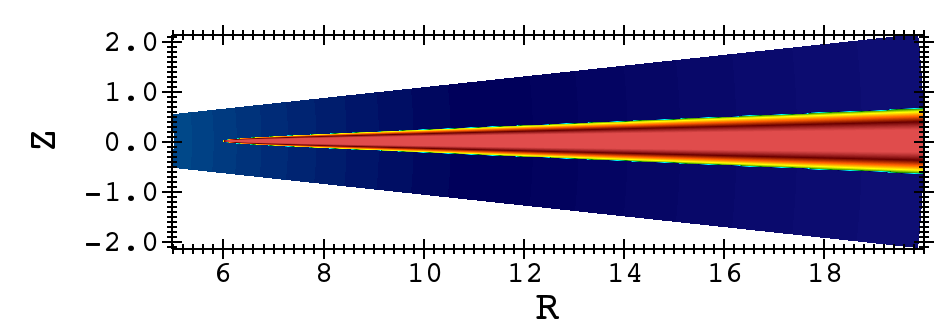}
\includegraphics[width=\columnwidth]{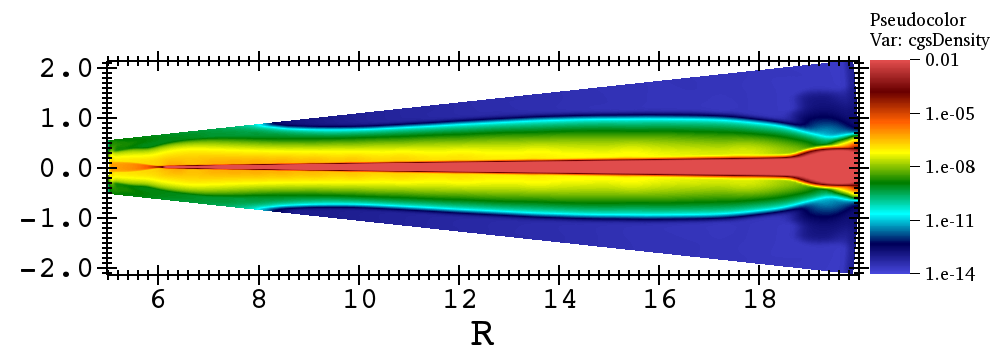}
\caption{Pseudocolor plots of disk mass density in $\mathrm{cgs}$ units. The left panel shows the initial setup and the right panel shows a snapshot at $t = 84000\,GM/c^3 \approx 910\,t_\mathrm{ISCO}$ (where, $t_\mathrm{ISCO} = 92.3\,GM/c^3$ is Keplerian orbital period at the ISCO). The horizontal and vertical axes are in units of $GM/c^2$.}
\label{disk}
\end{figure*}

The initial setup (left panel Figure~\ref{disk}) for the reported simulation is based on the standard thin disk model of \citet{Novikov73}, a relativistic generalization of the Shakura-Sunyaev alpha disk where accretion is driven by an assumed turbulent viscosity, with the effective viscous stress taken to be proportional to the pressure, the constant of proportionality being the $\alpha$-parameter \citep{Shakura73}.
It is important to realize that, unlike, e.g., the 3-D analytic \cite{KK00} solution of an alpha disk, the Novikov-Thorne disk is a solution to height-integrated equations, so it satisfies the 3-dimensional equations only approximately. The radiative transfer is also handled differently in the disk model and in our simulations.  Hence we do not expect a disk initialized with this solution to be in perfect equilibrium at the beginning of the simulation.

We initialize the disk as a gas pressure dominated setup, with the dominant opacity due to electron scattering. The motivation behind choosing a gas pressure dominated disk is to avoid the thermal and viscous instabilities present in radiation pressure dominated thin disks \citep{Lightman74, Shakura76, Piran78, Hirose09, Jiang13, Mishra16, Fragile2018}. 
 To get the initial vertical profile of the disk, we assume a vertically isothermal disk. The initial radial profiles are set for a constant initial mass accretion rate, $\dot{M} = 0.01\,L_\mathrm{Edd}/c^2$, where $L_\mathrm{Edd}$ is the Eddington luminosity. 
 
 The density and pressure profile at $t=0$ is then given by

\begin{equation}
    \rho(R,z) = \rho_\mathrm{NT}(R) e^{-z^2/2H^2},
\end{equation}

\begin{equation}
    p_\mathrm{tot}(R,z) = \frac{GM}{R^3}H^2\rho(R,z),
\end{equation}
where $\rho_\mathrm{NT}(R)$ and $H(R)$ are the radial profiles of midplane density and disk height, respectively, adopted from the \cite{Novikov73} disk model, $p_\mathrm{tot}=p_\mathrm{gas}+p_\mathrm{rad}$ is the total pressure, and $R$ is the cylindrical radius. 

%%+++++++++++++++++++++++++++++++++++++++++++++++++++++++++++++++++++++++++++++++++++++++++++++++++++++++++++++++++++++++++++++++++++++++++++++
The inner boundary of the simulation domain lies at $r_\mathrm{min} = 5\,GM/c^2$ and the outer boundary is at $r_\mathrm{max} = 20\,GM/c^2$. Note that we do not put the inner boundary at the Schwarzschild radius to save the computational time that would be required by the very small time-steps in the vicinity of event horizon ($r = 2GM/c^2$). The other motivation to have inner edge at $r = 5\,GM/c^2$, is that we are interested in the physics of accretion at or outside the innermost stable circular orbit (ISCO) to study disk oscillations. We use a logarithmic grid in the radial direction, $x_1 = 1 + \ln(r/r_\mathrm{BH})$, with $n_\mathrm{r} = 256$ shells, and a non-uniform polar grid along $\theta$ with $n_\theta = 192$ shells. The polar grid concentrates the highest resolution close to the disk mid-plane using the transformation
\begin{equation}
    \theta = x_2 + \frac{1}{2}[1 - q]\sin(2 x_2),
\end{equation}
where $q = 0.1$ and $x_2$ is a uniform polar coordinate. The polar domain spans the equatorial plane region with $\theta_\mathrm{min} - \theta_\mathrm{max} = 0.289\,\mathrm{rad}$, where $\theta_\mathrm{max}$ and $\theta_\mathrm{min}$ are polar boundaries of the domain, above and below the disk, respectively.

The boundary conditions are set to outflow condition along the inner radial and polar boundaries. The outer radial boundary assumes a constant boundary condition in which matter is constantly supplied to feed the disk with a mass accretion rate of $\dot{M} = 0.01\,L_\mathrm{Edd}/c^2$. The viscosity parameter $\alpha = 0.02$ is chosen to correspond to the effective stress in our previous global radiative GRMHD simulations \citep{Mishra16}. These values of mass accretion rate, $\dot{M}$, and viscosity parameter, $\alpha$, give a disk in stable equilibrium with surface mass density $\Sigma \approx 3\times 10^4\,\mathrm{g}\,\mathrm{cm}^{-3}$ and disk midplane temperature $T_\mathrm{c} \approx 10^7\,K$. Thus the initial profile of our numerical setup coincides with a local point on the stable gas-pressure-dominated branch of the $T-\Sigma$ thermal-equilibrium curve. 

The radiation transfer is treated using the M1 closure scheme \citep{Mihalas84,Sadowski13}. We consider a constant opacity due to electron scattering ($\chi_\mathrm{s} = 0.34\,\mathrm{cm}^2\,\mathrm{g}^{-1}$) together with a Kramer-style free-free absorption opacity. We also include thermal Comptonisation for interactions between electrons and photons. Further details of the simulation setup, viscosity prescription and GRRHD code used for this work are provided in \citet{Fragile2018}, where this simulation appears as S01E.
 
%%%%%%%%%%%%%%%%%%%%%%%%%%%%%%%%%%%%%%%%%%%%%%%%%%%%%%%%%
\section{Preliminary Analysis}
\label{outcome}
%%%%%%%%%%%%%%%%%%%%%%%%%%%%%%%%%%%%%%%%%%%%%%%%%%%%%%%
\subsection{Characteristic frequencies}
\label{eigenfreq}
%%%%%%%%%%%%%%%%%%%%%%%%%%%%%%%%%%%%%%%%%%%%%%%%%%%%%%%%%

%%+++++++++++++++++++++++++++++++++++++++++++++++++++++++++++++++++++++++++++++++++++++++++++++++++++++++++++++++++++++++++++++++++++++++++++++
\begin{figure}
\centering
\includegraphics[width=\columnwidth]{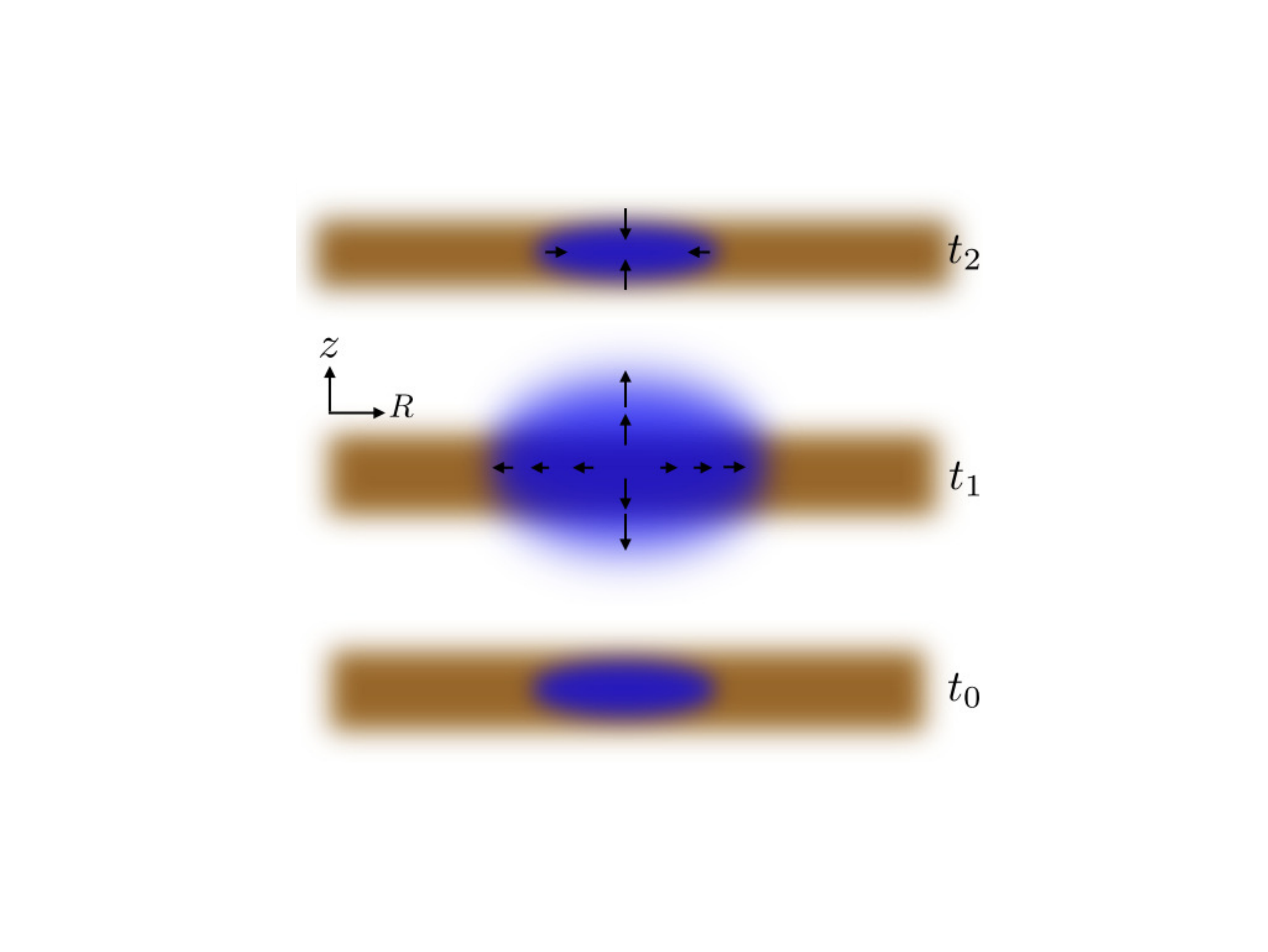}
\caption{Schematic diagram illustrating a breathing disk. The vertical axis shows time and the horizontal axis radius. The blue region shows a local breathing oscillation in the disk. The velocity vectors show the qualitative behavior of the breathing oscillation. For clarity, only an oscillation at a particular radius is shown. In our simulations the disk seems to execute such oscillations at all radii, each at its own frequency.}
\label{sketch}
\end{figure}
%%+++++++++++++++++++++++++++++++++++++++++++++++++++++++++++++++++++++++++++++++++++++++++++++++++++++++++++++++++++++++++++++++++++++++++++++

The vertical and radial epicyclic (angular) frequencies in the Schwarzschild metric are given by \citep[e.g.,][]{katoBook}
\begin{equation}
    2\mathrm{\pi}\nu_\perp (R)= \Omega(R) \, ,
    \label{freqs}
\end{equation}
%    \,\,\,
\begin{equation}
     \kappa(R) = \sqrt{1 - \frac{6\,GM}{Rc^2}}\,\,\Omega(R) \, ,
    \label{freqs2}
\end{equation}
where $\Omega(R) = \sqrt{GM/R^3}$ is the Keplerian orbital frequency. 

The epicyclic frequencies change with radial position in the disk, so oscillations at these, and related, frequencies do not correspond to global eigenmodes of the disk. Nevertheless, a useful guide to the expected local vertical motions can be obtained by considering the plane-parallel approximation, in which the radial variation of the frequencies is neglected.  

In the plane-parallel approximation of vertically polytropic disks, the velocity eigenfunctions (vertical component of the velocity as a function of $z$, the height above/below the midplane) of vertical oscillations are given by Gegenbauer polynomials, and the eigenfrequencies are given by the formula \citep{silb2001,katoBook,deepika2017}
\begin{equation}
    \nu_n = \sqrt{\frac{m(m+2n-1)}{2n}}\,\nu_\perp \, ,
    \label{eigenfreqs}
\end{equation}
with the polytropic index $\gamma=(n+1)/n$, and $m=1, 2, 3, ...$ the mode number. The breathing mode is the oscillation with a single node, in which the vertical component of the velocity is linearly proportional to $z$.  Note that the frequencies are not in a harmonic progression. Hence, any harmonics of the fundamental (breathing oscillation) seen in our PDS cannot correspond to higher order vertical oscillations, but must reflect nonlinearities in the oscillatory motion. A thin accretion disk around a black hole does not satisfy the plane-parallel approximation, so the oscillation is no longer a mode, as already remarked. However, we will be considering its local counterpart. The frequency of the breathing oscillation is given by \citep{deepika2017}
\begin{equation}
 \nu_\mathrm{b}\equiv\nu_2 = \sqrt{1 + \gamma}\,\nu_\perp.
     \label{bfreq}
\end{equation}
For vertical oscillations of our gas-pressure dominated disk, the polytropic index will be taken to be $\gamma=5/3$. 
As for this value of the polytropic index $\sqrt{1 + \gamma}\approx 1.633$ is very close in value to $5/3\approx 1.667$, it is important to exclude the possibility of a 5:3 resonance, entertained in the context of black hole QPOs  \citep[e.g.,][]{Kluzniak02}. We are confident that the PDS of our simulations have sufficient resolution to distinguish between 1.663 and 1.667, particularly for the simple harmonics of the fundamental (as seen in the top panels of Figure~\ref{pdsvr0}). However, we also provide eigenfunctions of the fluid variables to conclusively identify the reported frequencies with the breathing oscillation.

%%%%%%%%%%%%%%%%%%%%%%%%%%%%%%%%%%%%%%%%%%%%%%%%%%%%%%%%%
\subsection{Comparison with theoretical work}
\label{result0}
%%%%%%%%%%%%%%%%%%%%%%%%%%%%%%%%%%%%%%%%%%%%%%%%%%%%%%%%%
Our model evolves for $t = 84000\,GM/c^3$, which is much smaller than viscous time scale ($t_\mathrm{visc} = r^2/(\alpha c_\mathrm{s}H) \approx 10^9\,GM/c^3$ for $\alpha = 0.02$ and $r = 10$). This suggests that we have not achieved a steady state solution even over the inner one decade in radii. However, even over this shorter timescale, the disk relaxes from its initial Novikov-Thorne profile to a somewhat different state as shown in Figure~\ref{disk}.

%+++++++++++++++++++++++++++++++++++++++++++++++++++++++++++++++++++++++++++++++++++++++++++++++++++++++++++++++++++++++++++++++++++++++++++++
\begin{figure*}
\includegraphics[width=\columnwidth]{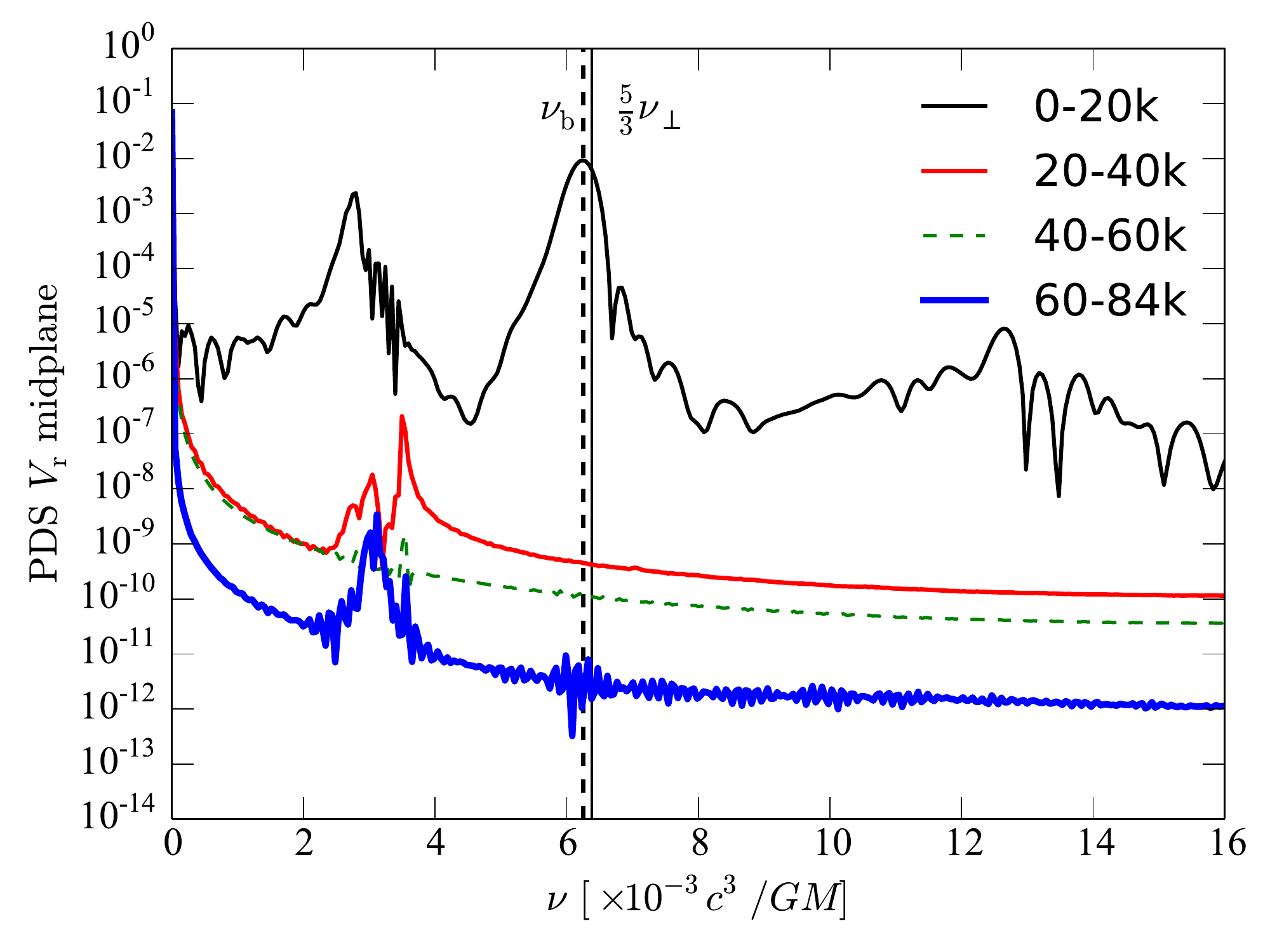}
\includegraphics[width=\columnwidth]{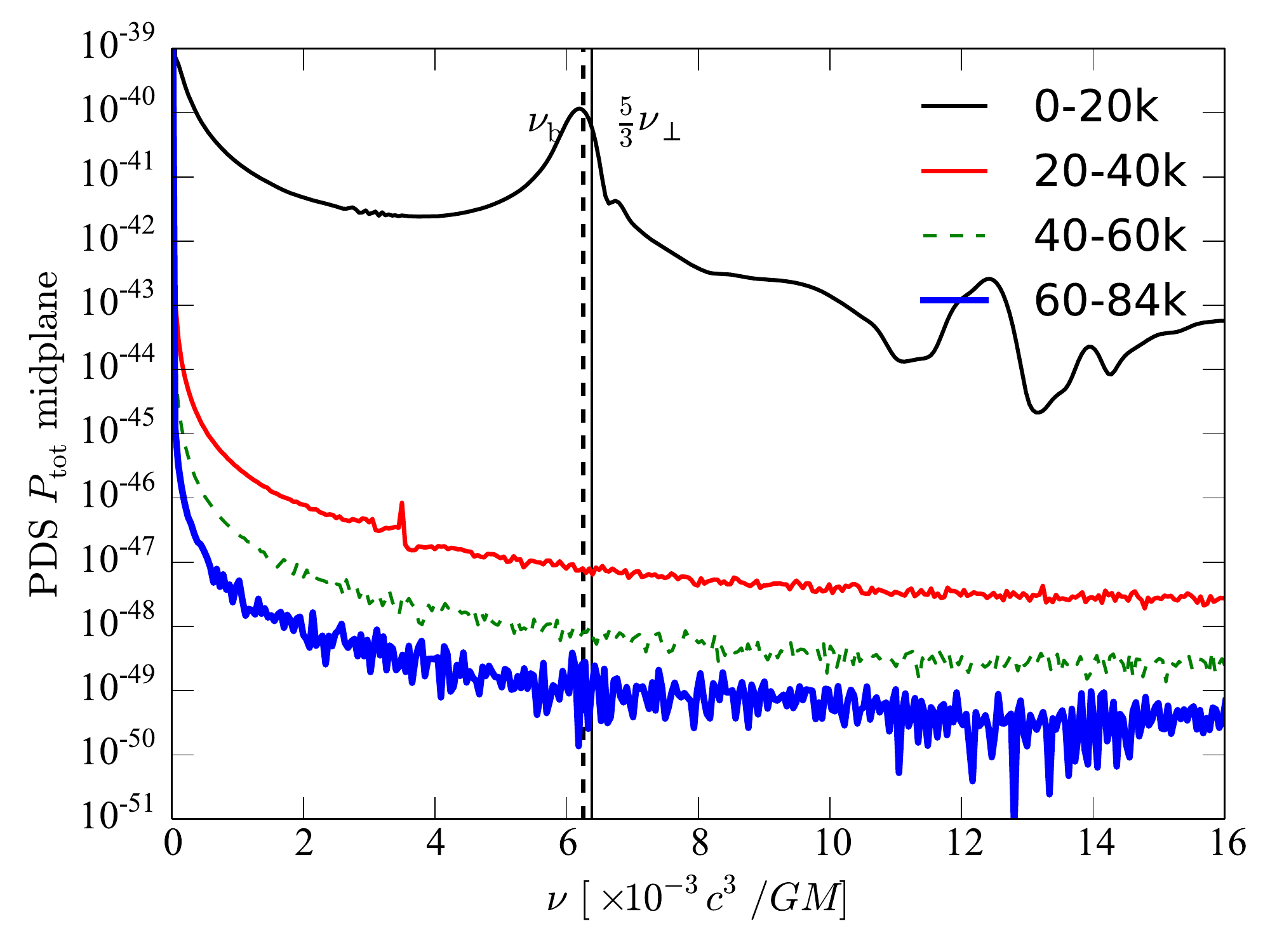}
\centering
\includegraphics[width=\columnwidth]{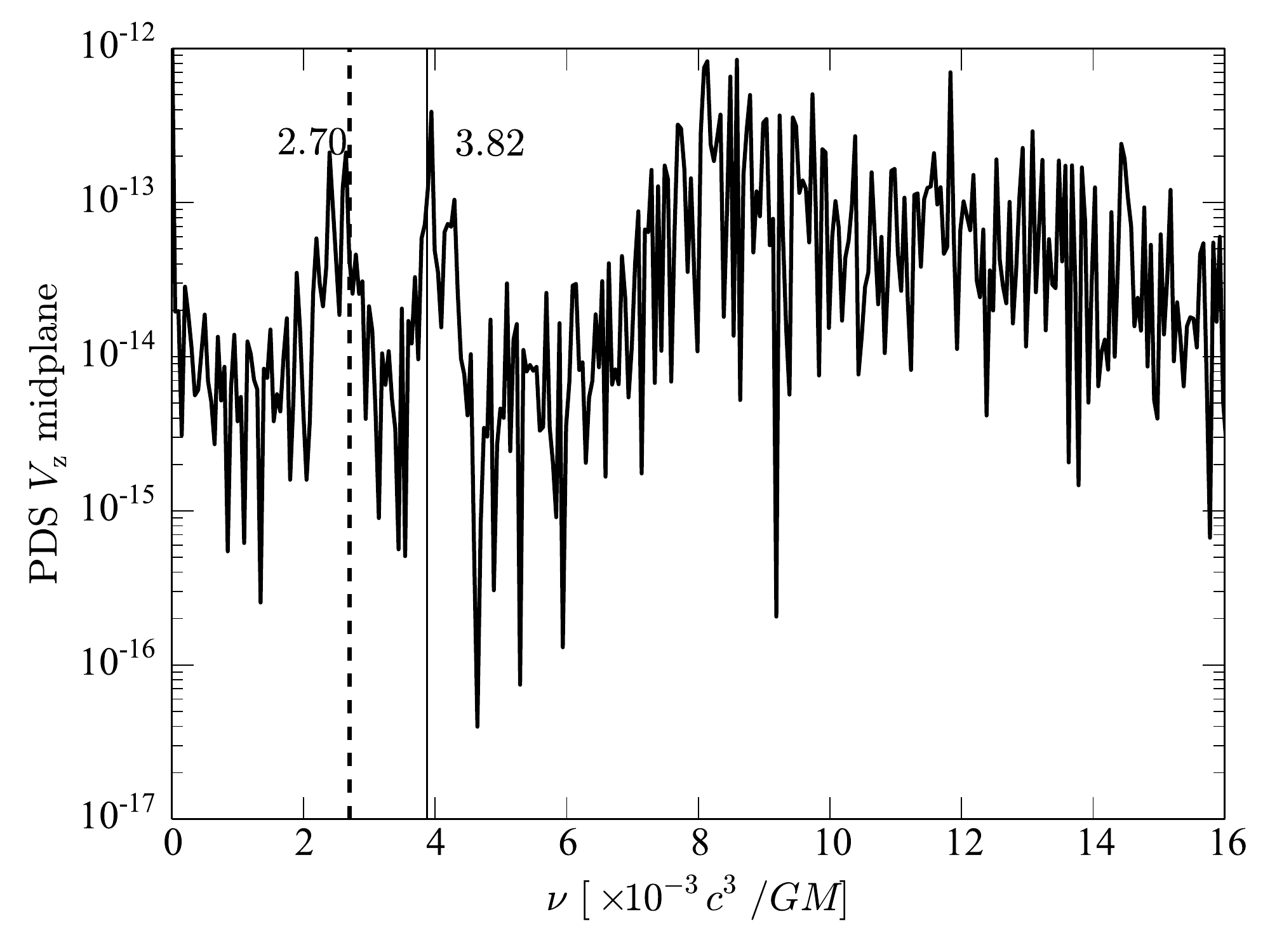}
\caption{Local PDS of midplane radial velocity (left panel), total midplane pressure (right panel), and midplane vertical velocity (lower panel) at $R = 12\,GM/c^2$. Different curves show PDS computed over four individual time windows. In the top two panels, the dashed vertical line shows breathing oscillation frequency, whereas the thin solid black vertical line shows $5/3\,\nu_\perp$, or 1.667 times the vertical epicyclic frequency. In the bottom panel, which only covers the initial $t = 20000 GM/c^3$, the dashed and solid vertical lines correspond to local radial and vertical epicyclic frequencies.}
\label{pdsr16}
\end{figure*}
%%+++++++++++++++++++++++++++++++++++++++++++++++++++++++++++++++++++++++++++++++++++++++++++++++++++++++++++++++++++++++++++++++++++++++++++++

Thus, even though we do not  intentionally introduce a perturbation to the disk structure, the adjustment of the disk to its steady state results in an initial oscillatory disturbance which includes a component with the velocity field shown schematically in Figure~\ref{sketch}. In Figure~\ref{pdsr16} we show that this motion temporarily induces peaks in the local power density spectra (PDS) of the midplane radial velocity (left panel) and the total midplane pressure (right panel). In addition to the expected local radial epicyclic frequency, a prominent peak at a higher frequency is seen in the PDS. At first glance it appear this may be at 5/3 the local vertical epicyclic frequency (lower panel PDS of midplane vertical velocity), which would be at $6.38\times10^{-3}c^3/(GM)$. However, as we will show, a closer examination reveals that this is actually an imprint of the breathing oscillation. A hint of this is already provided in the right panel of Figure~\ref{pdsr16}, showing the PDS of total pressure. 

Except for very extended structures, the simplest oscillatory motions of orbiting fluid are incompressible in character and occur at the vertical (perpendicular to the orbital plane) and radial epicyclic frequencies \citep{Kluzniak05}. Indeed,  as expected for incompressible motion, in spite of the prominent peak at the radial epicyclic frequency, at $2.7\times10^{-3}c^3/(GM)$, in the PDS of radial velocity (left panel of Figure~\ref{pdsr16}),  there is no corresponding peak in the PDS of the pressure (right panel of Figure~\ref{pdsr16}). However, the power maximum at the higher frequency of $\sim 6\times10^{-3}c^3/(GM)$ at early times (0-20k) is clearly visible in both the radial velocity and the total midplane pressure PDS, indicating that the motions at this higher frequency are associated with significant compression of the oscillating fluid. Also, quantitatively, the location of the power maximum in these PDS of velocity and pressure at an illustrative location (though this is true at all $R > 6$) in the disk, $R=12\,GM/c^2\equiv R_0$, precisely coincides with the frequency of the breathing oscillation at the same radius $\sqrt{1 + 5/3}\,\nu_\perp(R_0) = 6.25\times 10^{-3}\,c^3/GM$, c.f. Eqs.~(\ref{freqs}) and (\ref{bfreq}). 
In lower panel of Figure~\ref{pdsr16}, we show local PDS for midplane vertical velocity at $R = 12\,GM/c^2$. Since, we mainly focus on first $t = 20 000\,GM/c^3$, the PDS  is computed only for this time window. There are three prominent peaks in this PDS. The dashed and solid vertical lines correspond to local radial and vertical epicyclic frequencies. The third peak correspond to third harmonic of radial epicyclic frequency.

%%%%%%%%%%%%%%%%%%%%%%%%%%%%%%%%%%%%%%%%%%%%%%%%%%%%%%%%%%
\section{Breathing Oscillations}
\label{results}
%%%%%%%%%%%%%%%%%%%%%%%%%%%%%%%%%%%%%%%%%%%%%%%%%%%%%%%%%%
For the remainder of the paper, we report the disk oscillations using data for the first $20000\,GM/c^3\approx 216\,t_\mathrm{ISCO}$ of the simulation, and all time averages are over this same interval. We preface our discussion of the character of the inferred oscillations with a discussion of the diagnostics we use to probe them. These will allow us to construct and interpret power density spectra of the relevant quantities, and to extract their flow pattern. All midplane quantities (radial and vertical components of the three velocity and total pressure) are computed by averaging over the two cells straddling the equatorial plane at each radius.

\bigskip
%%%%%%%%%%%%%%%%%%%%%%%%%%%%%%%%%%%%%%%%%%%%%%%%%%%%%%%%%
\subsection{Diagnostics}
\label{diagn}

 In the breathing oscillation, the vertical expansion of the disk is symmetric on both sides of its midplane (Figure~\ref{sketch}). For this reason we will examine the PDS of the vertical component of the velocity averaged over one half of the domain, from $\theta_\mathrm{max}$ to $\mathrm{\pi}/2$.
The vertical velocity in the upper half of the disk is computed as a density weighted, vertical cell average in the following way. 
\begin{equation}
    <V_\mathrm{z}(R)>_\rho = \frac{\int^{\pi/2}_{\theta_\mathrm{max}}\sqrt{-g}\rho(R,\theta) V_\mathrm{z}(R,\theta)d\theta}{\int^{\pi/2}_{\theta_\mathrm{max}}\sqrt{-g}\rho(R,\theta)d\theta},
\end{equation}
where $\rho$ is mass density in the coordinate fixed frame and $g$ is metric determinant. The cooling rate is computed at the photosphere of the disk by measuring net vertical component of radiation flux,
\begin{equation}
    Q^-(R) = F^z_\mathrm{photo+}(R) - F^z_\mathrm{photo-}(R),
\end{equation}
where $F^z_\mathrm{photo} = -4/3E_R u^z_R (u_R)_t$. The $E_R$ and $u_R$ correspond to the radiation energy density in the radiation rest frame and the four velocity of radiation rest-frame, respectively. Subscripts ``$\mathrm{photo+/-}$'' correspond to the photosphere above and below the equator, respectively. We define the photosphere as the surface with optical depth unity $(\tau = 1)$. In order to compute the optical depth, we integrate the total opacity ($\chi\rho$ for $\chi = \chi_\mathrm{s} + \chi_\mathrm{a}$, where $\chi_\mathrm{s}$ and $\chi_\mathrm{a}$ are opacity coefficients due to electron scattering and absorption) from $\theta_\mathrm{min}$ or $\theta_\mathrm{max}$ to the point when it reaches unity, as 
\begin{equation}
    \tau_\mathrm{min}(R,\theta^\prime) = \int^{\theta^\prime}_{\theta_\mathrm{min}}u^t\chi\rho(R,\theta)\sqrt{g_{\theta\theta}}d\theta,
\end{equation}
or
\begin{equation}
    \tau_\mathrm{max}(R,\theta^\prime) = -\int_{\theta_\mathrm{max}}^{\theta^\prime} u^t\chi\rho(R,\theta)\sqrt{g_{\theta\theta}} d\theta.
\end{equation}

\begin{figure*}
\centering
\includegraphics[width=2\columnwidth]{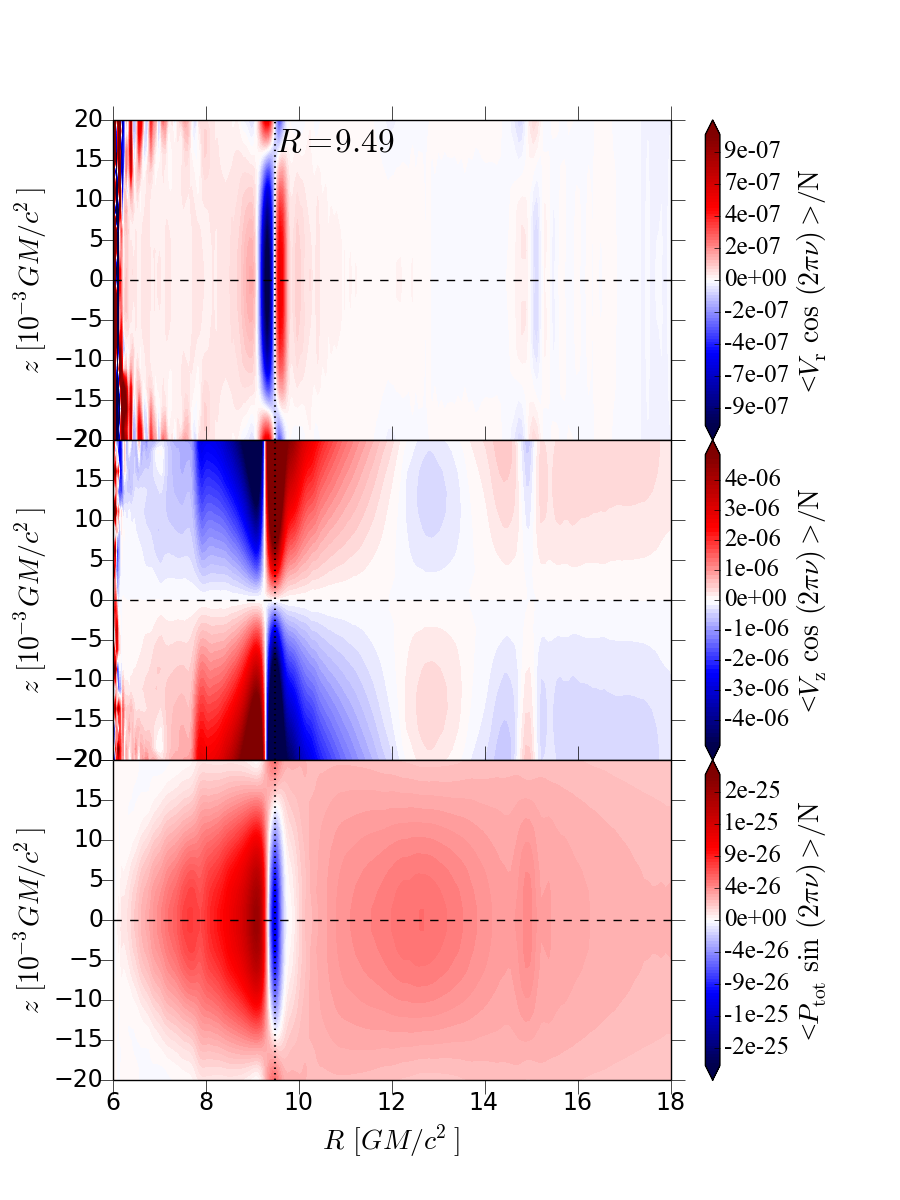}
\caption{Eigenfunctions of the radial velocity (top), vertical velocity (middle) and total pressure (bottom). The frequency $\nu \approx 9\times 10^{-3}\,c^3/GM$ is used for all three panels. The vertical dotted line shows the location of the radial node of breathing oscillations at this frequency. The horizontal dashed line shows the disk midplane. The plots are normalized by the number of oscillation cycles $N = \nu T$.}
\label{efunc}
\end{figure*}

The eigenfunctions are computed from the time integral of the chosen fluid field times the appropriate trigonometric function, at a specific frequency ($\nu^R_\mathrm{b} $):
\begin{equation}
    <f(R,z)>_t = \int^{t_2}_{t_1} f(R,z)\cos(2\pi\nu^R_\mathrm{b} t)dt
    ~\mathrm{or}~ \int^{t_2}_{t_1} f(R,z)\sin(2\pi\nu^R_\mathrm{b} t)dt,
    \label{eigenfun}
\end{equation}
where $f$ could be the radial or vertical component of the fluid velocity or the total midplane pressure. We search for power in both the real and imaginary components using the cosine and sine, respectively. In Section \ref{efunc}, we report the eigenfunctions for frequency $\nu^R_\mathrm{b} = 9\times 10^{-3}\,c^3/GM$, corresponding by Eq.~\ref{bfreq} to the breathing oscillation frequency at $R = 9.49\,GM/c^2$.

%%+++++++++++++++++++++++++++++++++++++++++++++++++++++++++++++++++++++++++++++++++++++++++++++++++++++++++++++++++++++++++++++++++++++++++++++
\begin{figure*}
\includegraphics[width=\columnwidth]{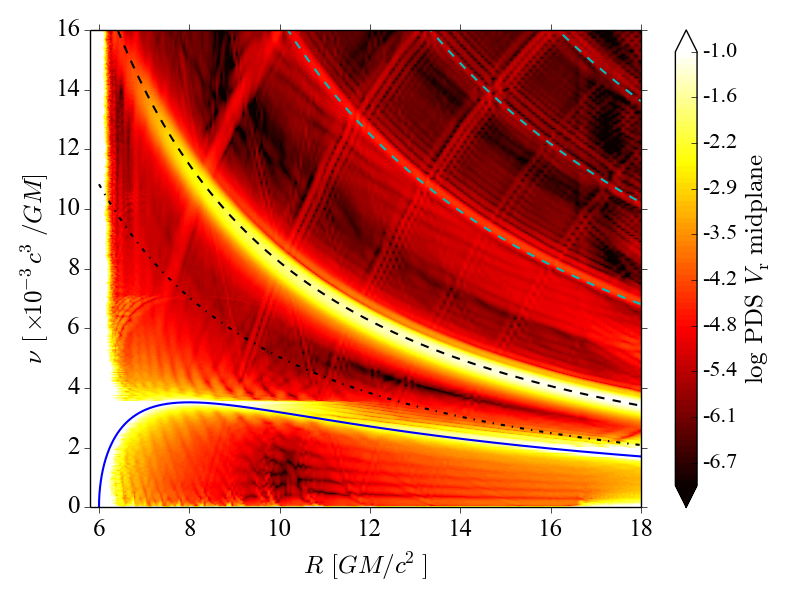}
\includegraphics[width=\columnwidth]{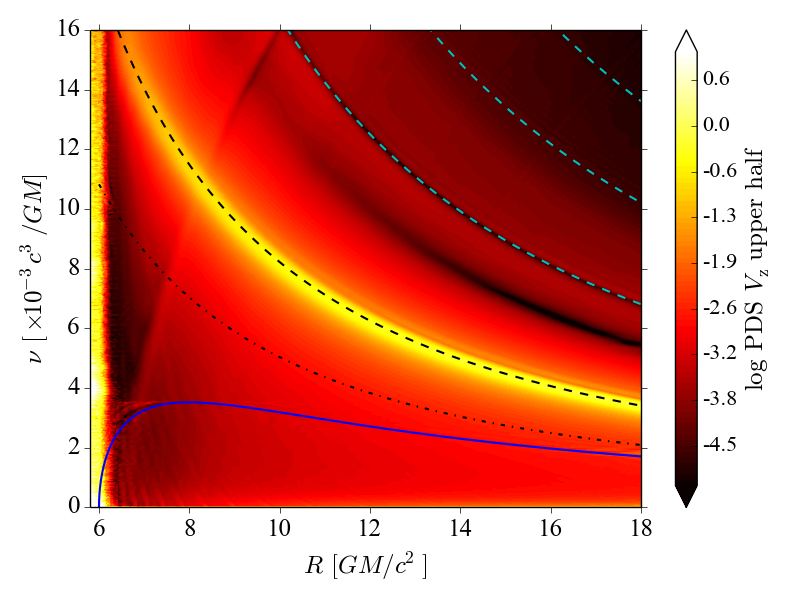}
\centering
\includegraphics[width=\columnwidth]{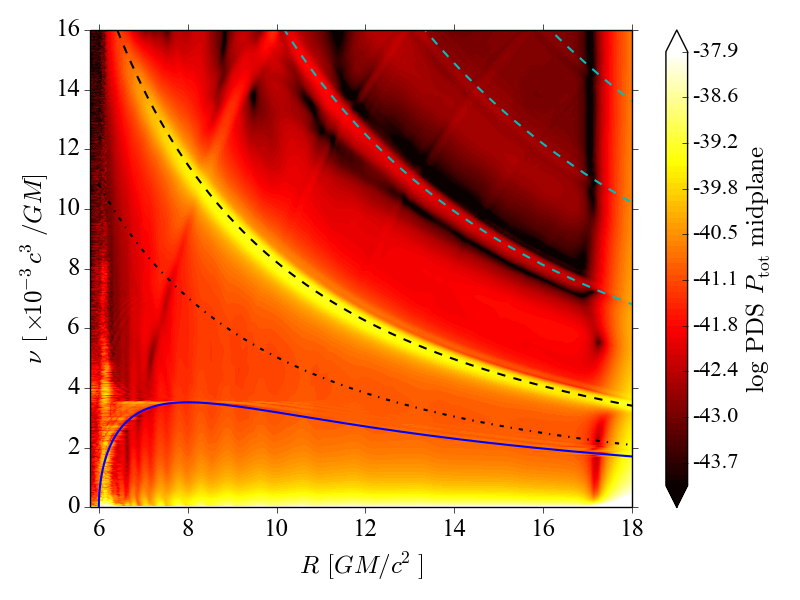}
\includegraphics[width=\columnwidth]{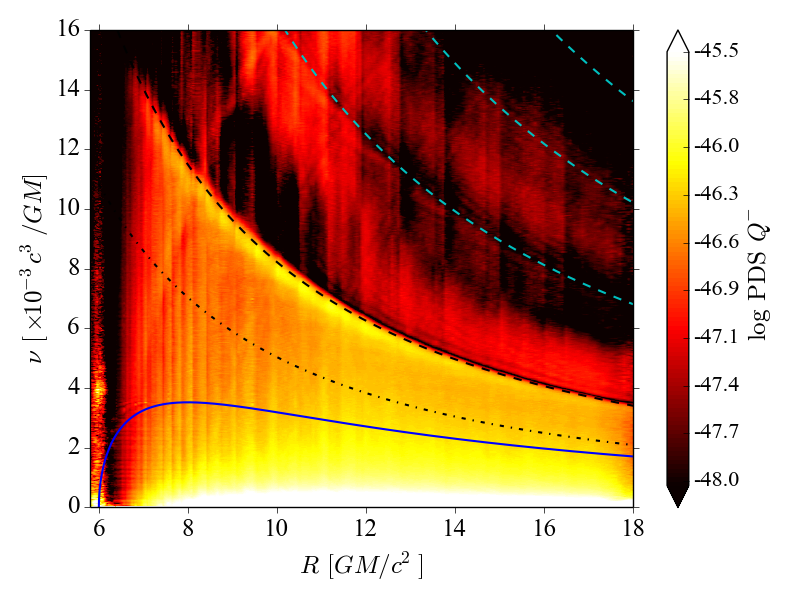}
\caption{Radial profiles of the PDS of the midplane radial fluid three velocity (upper left); shell averaged vertical velocity above the midplane (upper right); total midplane pressure (lower left); and radiative cooling rate computed at the photosphere (lower right). The solid blue curves correspond to the radial epicyclic frequency, $\kappa/(2\mathrm{\pi})$, dotted-dashed curves correspond to the vertical epicyclic frequency, $\nu_\perp$, Eq.~(\ref{freqs}), and the dashed curves correspond to the breathing oscillation frequency and its higher harmonics. The stripes of power curving up to the right are aliases of the high-frequency (low-radius) extension of the curves of enhanced power. }
\label{pdsvr0}
\end{figure*}
%%+++++++++++++++++++++++++++++++++++++++++++++++++++++++++++++++++++++++++++++++++++++++++++++++++++++++++++++++++++++++++++++++++++++++++++++

%%%%%%%%%%%%%%%%%%%%%%%%%%%%%%%%%%%%%%%%%%%%%%%%%%%%%%%
%%%%%%%%%%%%%%%%%%%%%%%%%%%%%%%%%%%%%%%%%%%%%%%%%%%%%%%
\subsection{Power density spectra}
\label{breath}
The power density spectra in Figure~\ref{pdsvr0} were computed by performing fast-Fourier transforms (FFT) of the time-series of midplane variables (radial velocity and total pressure), vertical velocity in the upper half of the simulation domain, and cooling rate.
The PDS of the midplane radial component of fluid three velocity (upper left panel, Figure~\ref{pdsvr0}) shows two high power and three moderate power bands. The power concentrated along the solid blue curve demonstrates radial oscillations at the radial epicyclic frequency. The second high power band follows the dashed curve corresponding to the breathing mode frequency of Eq.~(\ref{bfreq}) for a polytropic index $\gamma = 5/3$, i.e. $\nu_\mathrm{b} = \sqrt{8/3}\,\nu_\perp$. 

The vertical velocity PDS shows the first direct evidence for breathing oscillations, apart from the frequency. In order to understand the properties of these breathing oscillations more thoroughly, we show PDS of the vertical component of fluid velocity in the upper half of the disk (upper right panel, Figure~\ref{pdsvr0}). We see a band with high power following the breathing oscillation frequency, which is absent in the vertical velocity PDS of the complete vertical domain (not shown). This confirms the presence of a reflection symmetric, volume changing (compression/rarefaction) vertical motion. The periodic changes in the specific volume lead to variations in pressure at the same frequency, as shown in the lower left panel of Figure~\ref{pdsvr0} for the total midplane pressure.  The first harmonic of the breathing oscillation can also be seen in this PDS. This gives threefold evidence for the presence of breathing oscillations: vertical motion symmetric with respect to reflection in the midplane of the disk, and direct evidence of pressure variations, both at the frequency given by Eq.~(\ref{bfreq}). Interestingly, no power appears at the radial epicyclic frequency in PDS of the total midplane pressure.  Again, this illustrates the compressible and incompressible natures of the breathing and radial epicyclic oscillations, respectively.

The higher frequency power bands seen in the global PDS of the midplane radial velocity and the total midplane pressure (Figure~\ref{pdsvr0}) and the higher frequency peaks in the local PDS (Figure~\ref{pdsr16}) are harmonics ($2\nu_\mathrm{b}$, $3\nu_\mathrm{b}$ and $4\nu_\mathrm{b}$) of the breathing oscillation frequency, implying nonlinearity of the oscillation as, clearly, the power is not limited to one harmonic component. 

The simulation includes radiative transfer self-consistently (hydrodynamics coupled with radiation); hence we can also look for modulations of the emitted flux from the disk. We compute the vertical radiative flux at the photosphere of the disk to estimate the cooling rate. The lower right panel in Figure~\ref{pdsvr0}, shows the PDS of the cooling rate. We do notice some power enhancement along the curve following the breathing oscillation frequency but not much in its harmonics (the first harmonic, most clearly seen in its alias at $R<10$, is barely visible). This could be due to the harmonics of the breathing oscillations being too weak to be seen in the variability of the radiative flux at the photosphere, perhaps because they are mainly concentrated close to the midplane of the disk.  However, the radiative coupling of the oscillations will require a more careful investigation.

%%%%%%%%%%%%%%%%%%%%%%%%%%%%%%%%%%%%%%%%%%%%%%%%%%%%%%
%%%%%%%%%%%%%%%%%%%%%%%%%%%%%%%%%%%%%%%%%%%%%%%%%%%%%%
%%+++++++++++++++++++++++++++++++++++++++++++++++++++++++++++++++++++++++++++++++++++++++++++++++++++++++++++++++++++++++++++++++++++++++++++++
%%+++++++++++++++++++++++++++++++++++++++++++++++++++++++++++++++++++++++++++++++++++++++++++++++++++++++++++++++++++++++++++++++++++++++++++++

\subsection{Eigenfunctions}
\label{eigenfunc}
The breathing oscillations can be further investigated by studying their eigenfunctions, which quantify the flow patterns of the oscillations. In Figure~\ref{efunc}, we present the eigenfunctions of the radial velocity, vertical velocity and total midplane pressure for $\nu^R_\mathrm{b} \approx 9\times 10^{-3}\,c^3/GM\equiv \nu$, where the superscript $R$ reminds us that we are considering a specific radius, in this case $R = 9.49\,GM/c^2$. In general, one can choose any frequency up to the maximum at ISCO. Since these oscillations are present at all radii, a different frequency will simply pick the flow pattern eigenfunction at a correspondingly different radius. The eigenfunctions are normalized by $N = \nu T$, where $T = 20000\,GM/c^3$ is the time window over which the integral was performed. 

The top panel in Figure~\ref{efunc} exhibits the meridional pattern of radial motion (at frequency  $\nu\approx 9\times 10^{-3}\,c^3/GM$), and shows a sequence of fairly narrow blue, white and red vertical stripes (the colors corresponding to negative, zero, and positive values, respectively). The thin white, vertical stripe, highlighted by the black dotted vertical line shows the (radial) node of the breathing oscillation, which indeed is located at the expected radius of $R\approx 9.49\,GM/c^2$. The vertical axis  approximately spans the disk height. The radial component of velocity flips spatial phase above $z\approx 0.015\,GM/c^2$. This could be a boundary effect; note that the oscillatory pressure variations seen in the total pressure eigenfunction plot (bottom panel) do not extend this high in the disk. 

The middle panel shows the eigenfunction of the vertical velocity component. The black dotted vertical line at $R= 9.49\,GM/c^2$ indicates a region of 
%overlaps with a
strong breathing motion. The flow is moving away from the disk midplane  (at this phase of the oscillation), with the vertical velocity increasing away from the midplane. This is a clear hallmark of the breathing oscillation and leads to the pressure decrease seen in the bottom panel. The rarefaction is enhanced by radial motion, with radial outflow from the region of the vertical oscillation occurring in phase with the vertical one---note the line of nodes in the radial velocity (top panel, at $R= 9.49\,GM/c^2$) is coincident with the radial location of maximum vertical velocity.
It would appear that the inner disk responds in turn to this drop in pressure with a counterflow. This is clearly shown in all panels of Figure~\ref{efunc}, with another sharp line of nodes in the radial motion, at $R= 9.1\,GM/c^2$, corresponding to a maximum of compression, as seen in the pressure plot in the lower panel, and coincident with another maximum in the perturbation of the vertical component of velocity, in opposite phase to the one at $R= 9.49\,GM/c^2$.
Thus,  to the left of the white vertical line in the middle panel we see vertical breathing motion of the same frequency, but opposite phase (compressive motion) relative to the oscillation at $R= 9.49\,GM/c^2$. No such counterflow is visible at larger radii, where the perturbation in $V_r$ tapers off gently.

An inspection  of the PDS in Figure~\ref{pdsvr0} reveals that in addition to the breathing mode at $R = 9.49\,GM/c^2$, the frequency $\nu \approx 9\times 10^{-3}\,c^3/GM$ also corresponds to the first harmonic of the breathing oscillation at $R\approx 14.9\,GM/c^2$. This harmonic is clearly seen in the eigenfunctions in  Figure~\ref{efunc} at $R\approx 14.9\,GM/c^2$. This feature is a carbon copy of the breathing oscillation eigenfunctions at $R = 9.49\,GM/c^2$, except for a (random) phase difference leading to a reversal of the sign (blue is mapped into red and vice versa in the color coding of the figure). 

Since the breathing oscillations are compressible, they are also seen in PDS of the total midplane pressure. In the bottom panel of Figure~\ref{efunc}, the eigenfunction of the total pressure shows radial structure compatible with the velocity field depicted in the top and middle panels. Strong vertical oscillations seen in the middle panel cause a decrease in pressure in the region in which flow is moving away from the midplane of the disk at $R\approx 9.49\,GM/c^2$, whereas the strong (positive) power to the left of the white vertical stripe is caused by compression (vertical flow towards the midplane at $R\approx9.1\,GM/c^2$, as seen in the middle panel). 

Note that the radial and vertical velocity eigenfunctions are depicted using the cosine component, whereas the total pressure eigenfunction uses the sine component. This shows the expected phase lag of $\mathrm{\pi}/2$ of the vertical component of fluid velocity behind the total pressure. Indeed, we obtain in the complex-number representation of polytropic vertical oscillations that the pressure perturbation leads the vertical velocity by a factor of the imaginary unit $\mathrm{i} = \sqrt{-1}$\,,
corresponding to a  $\mathrm{\pi}/2$ phase difference. Using the equation of continuity close to the disk midplane,
%[TO BE CHECKED CAREFULLY]. 
\begin{equation}
    \frac{\partial\rho}{\partial t} + \nabla.(\rho \vec{V}) = 0,
\end{equation}
where $\vec{V}$ is fluid velocity three-vector,
and substituting perturbations in density and the vertical component of the fluid three-velocity of the form $\delta \rho \exp (\mathrm{i}\omega t)$ and $V_z\exp (\mathrm{i}\omega t)$, with frequency $\nu=\omega/(2\mathrm{\pi})$,
one finds
\begin{equation}
    -\mathrm{i}\omega\delta\rho = \rho_0\frac{\partial V_\mathrm{z}}{\partial z},
\end{equation}
and hence, with $P_0 = K\rho_0^\gamma$ the equilibrium value of pressure and the perturbations in pressure consequently satisfying $\delta P/P_0=\gamma\delta\rho/\rho_0$, we obtain the phase lag from
\begin{equation}
    \frac{\delta P}{P_0} = \mathrm{i}\left(\frac{\gamma}{\omega}\frac{\partial V_\mathrm{z}}{\partial z}\right).
\end{equation}
 For clarity, we have omitted in this derivation the radial velocity contribution to the divergence, which in our simulation is in phase with the one for the vertical component.

%%%%%%%%%%%%%%%%%%%%%%%%%%%%%%%%%%%%%%%%%%%%%%%%%%%%%%

\section{Discussion}
\label{discuss}
We have found an example of a geometrically thin disk, initialized on the stable branch of thermal equilibrium curve, that exhibits breathing oscillations. While we do not intentionally perturb our initial setup, there is an unavoidable initial adjustment of the 
disk to its true equilibrium state, during which we see the breathing oscillations arise. We find that the amplitude of the oscillations diminishes after $t > 3000\,GM/c^3$, suggesting that any QPOs associated with breathing oscillations may only be seen around the times of X-ray spectral state transitions (or other significant perturbations to the disk). 

It is thought that in the low, hard spectral state the disk structure changes at a certain radius ({\it transition radius}) from a cold, thin disk to a hotter, thicker flow. This radius, a few times the ISCO one, is associated with particular frequencies, e.g., the orbital, radial and vertical (same as orbital frequency for a non-rotating black hole) epicyclic frequencies at that radius. If, during the spectral state transition, the inner part of the thin disk is perturbed more strongly near the {\it transition radius} than at larger radii, then this could lead to breathing oscillations reaching observable amplitudes near a specific frequency. The inferred frequency of such breathing oscillations in our simulation is in agreement with theory \citep{silb2001,katoBook,deepika2017}. In the following sections, we shall report two astrophysically relevant aspects of our work. 

\subsection{Implications for Variable X-ray Binaries}
 Observational studies of GRS 1915+105 have shown that it makes frequent transitions from a high, soft (disk) state to a corona-dominated (hard or no-disk) state and vice-versa \citep{Belloni2000}. Such a transition in an accreting system could lead to a triggering of the breathing and vertical oscillations. Also, GRS 1915+105 is a highly variable accreting black hole binary system, exhibiting a pair of oscillation frequencies with an approximate ratio of 5:3. We stated earlier that the ratio of the breathing oscillation frequency to the vertical epicyclic frequency is 1.63 (Eq.~\ref{bfreq}), which is close to 5:3 \citep[e.g.][]{Kluzniak02}. In order to connect our study with observations, we suggest a link between the breathing and vertical oscillations seen in our simulation and the QPOs in GRS 1915+105. However, in order for this to be the case, the breathing and vertical oscillations inferred in our numerical model need to appreciably modulate the emitted X-rays near a particular radius. Looking at lower right panel of Fig.~\ref{freqs}, we do see modulation (at all breathing oscillation frequencies) in the emitted radiative flux from the disk photosphere. The vertical oscillations, being incompressible, do not directly imprint their modulation on the PDS of radiative flux. However, geometric effects not taken into account in this analysis might provide a way for the vertical epicyclic frequency to also modulate the light curve \citep{Bursa04,Mishra17}. Under such favorable conditions, one might expect to observe high frequency QPOs not dissimilar to the simultaneous 69.2 Hz and 41.5 Hz pair reported in GRS 1915+105. 
%Additionally, such a pair of breathing and vertical oscillations seem to be very short lived in our simulations, which could also be true in the real astrophysical accreting systems because, observationally, QPOs are not yet seen in X-ray spectra of the high, soft state.

%%+++++++++++++++++++++++++++++++++++++++++++++++++++++++++++++++++++++++++++++++++++++++++++++++++++++++++++++++++++++++++++++++++++++++++++++
 \begin{figure}
 \centering
 \includegraphics[width=\columnwidth]{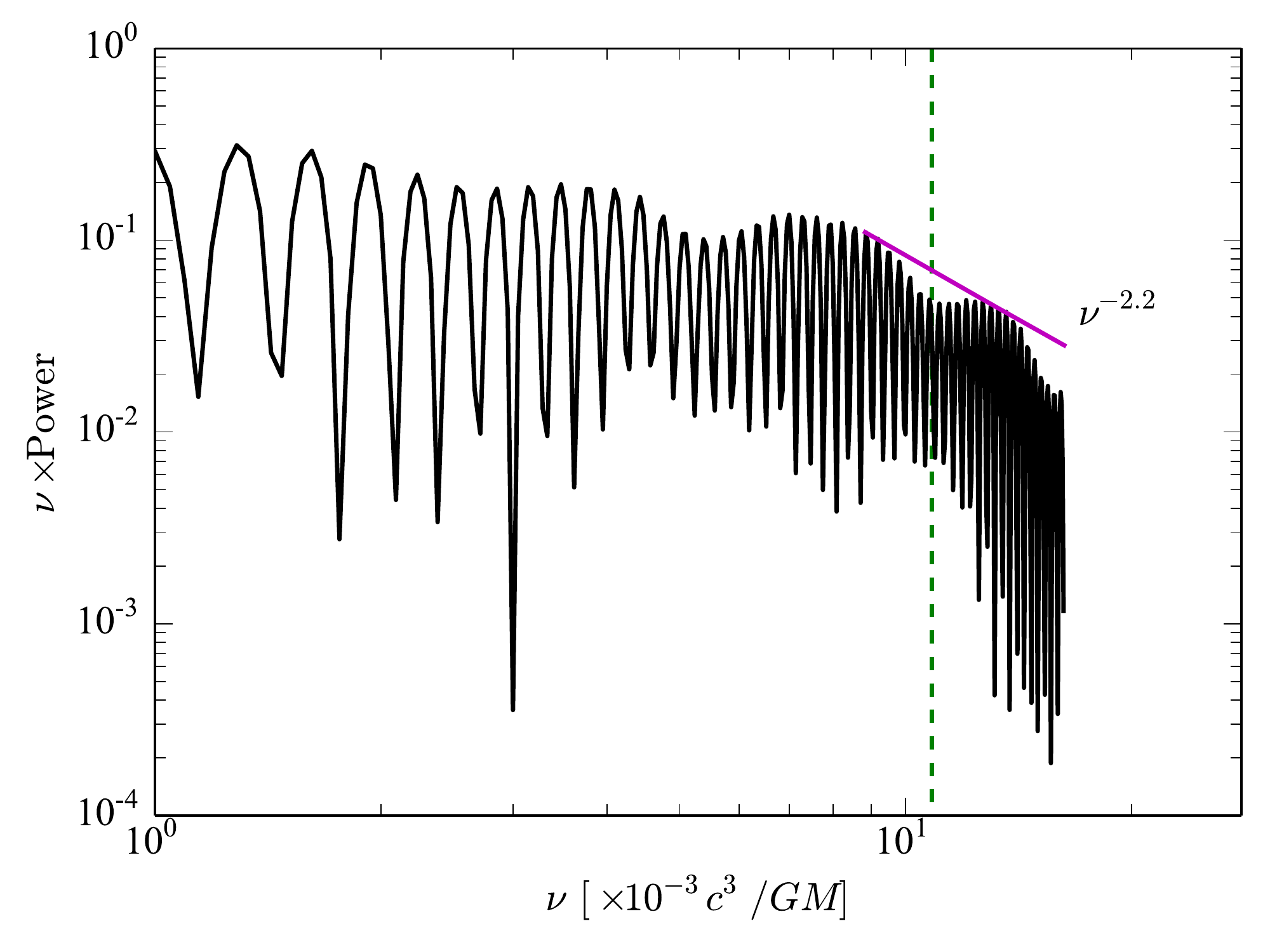}
 
     \caption{PDS of the lightcurve in the beginning ($t<20\,000\,GM/c^3$) of the simulation. Note that the steep, high-frequency, red noise clearly extends well beyond the ISCO frequency of  $\nu_\mathrm{ISCO}\approx 10.82\times10^{-3}c^3/(GM)$ (vertical, dashed, green line). This is because breathing oscillations occur at supra-orbital frequencies. The purple line traces a slope of $-2.2$ for comparison.}
     \label{rednoise}
\end{figure}

\subsection{Supra-orbital Frequencies}
It has long been thought that the highest frequency of X-ray variability corresponds to the orbital frequency at the ISCO, whether in the form of flickering, such as observed in Cyg X-1 \citep{Meekins84}, or kHz QPOs, with implications for the mass of the source inferred from the observed frequencies \citep{Kluzniak90,Kluzniak1998}. While this may be true for a thin accretion disk in its quiet, thermal (high, soft) state, our results clearly show that a perturbed disk may exhibit higher frequencies. As a first step towards converting the internal motions of the disk displayed in the PDS of Figure~\ref{pdsvr0} to an observable quantity, we plot the PDS of the lightcurve in the beginning interval, $0-20\,000\,GM/c^3$, of the simulation. At high frequencies, red noise is seen which is steeper than the $\nu^{-2}$ noise reported for sandpiles \citep{Jensen1989}.
 The red noise (i.e., variability) clearly extends past the maximum (ISCO) orbital frequency in the disk, $\nu_\mathrm{ISCO}\approx 10.82\times10^{-3}c^3/(GM)$ (dashed, vertical green line in Figure~\ref{rednoise}).

\section{Conclusions}
\label{conclusions}
We further investigated the stable gas-pressure-dominated, geometrically thin disk simulation reported in \citet{Fragile2018} to understand the time variability in it. These simulations could have astrophysical interest in the context of high frequency QPOs in black hole, low-mass X-ray binaries. Our simulation shows local oscillations at the breathing frequency predicted theoretically \citep{silb2001,deepika2017}. A qualitative, as well as quantitative, analysis of the simulation confirms the identity of the oscillation. The summary of results in this work is as follows:
\begin{itemize}
\item[1.] Breathing oscillations were identified for the first time in global numerical simulations of an accretion disk. Previously they had only been seen in a shearing box simulation \citep{Blaes2011}, which by the nature of its periodic boundary conditions cannot yield information on the radial dependence of fluid variables.
\item[2.] The power density spectra of the radial component of the fluid velocity and the total midplane pressure show large power at all radii at a specific frequency equal to the local breathing oscillation at each radius. In addition we see the first three harmonics of this oscillation.
\item[3.] The PDS for the vertical velocity component above the midplane of the disk shows the same high-power band as in the radial velocity and total midplane pressure PDS. However, the PDS for the vertical velocity component averaged over the entire polar range of the simulation domain does not show show excess power at the breathing oscillation frequency. This confirms that the disk is exhibiting breathing motion (expanding away from and contracting toward the midplane).
\item[4.] The PDS of the cooling rate of the disk also shows variability at the breathing oscillation frequency. This suggests that these oscillations are sufficiently strong to modulate the emitted radiation flux from geometrically thin disks.
\item[5.] 
For the first time, we showed the radial velocity eigenfunction for the breathing oscillation. This eigenfunction was not available previously, as the analytic calculations had been carried out in the plane parallel approximation (i.e. neglecting motions parallel to the disk midplane). 
\item[6.] Additionally, we showed the eigenfunctions of the vertical velocity component and the total pressure. The flow patterns in both show the expected pattern for breathing oscillations.
\item[7.]
The breathing and vertical oscillations in the disk could explain the 69.2 Hz and 41.5 Hz QPOs seen in X-ray observations of GRS 1915+105.

\end{itemize}

%%%%%%%%%%%%%%%%%%%%%%%%%%%%%%%%%%%%%%%%%%%%%%%%%%%%%%%%%

\section*{Acknowledgements}
BM thanks Omer Blaes for fruitful discussions. We acknowledge computational support from the PROMETHEUS supercomputer in the PL-Grid infrastructure in Poland, and the Extreme Science and Engineering Discovery Environment (XSEDE), which is supported by National Science Foundation grant number ACI-1053575. BM acknowledges support from NASA Astrophysics Theory Program grants NNX16AI40G and NNX17AK55G. WK and BM acknowledge support from the Polish NCN grant 2013/08/A/ST9/00795. This research was supported in part by the National Science Foundation under grants NSF PHY-1125915 and NSF AST-1616185. We also acknowledge the Kavli Institute for Theoretical Physics (KITP) for hosting us while this project was ongoing.
\\
Cosmos++ \citep{Anninos05,Fragile12,Fragile14}

\bibliographystyle{mnras}
\bibliography{ref} % if your bibtex file is called example.bib

\begin{thebibliography}{}
\makeatletter
\relax
\def\mn@urlcharsother{\let\do\@makeother \do\$\do\&\do\#\do\^\do\_\do\%\do\~}
\def\mn@doi{\begingroup\mn@urlcharsother \@ifnextchar [ {\mn@doi@}
  {\mn@doi@[]}}
\def\mn@doi@[#1]#2{\def\@tempa{#1}\ifx\@tempa\@empty \href
  {http://dx.doi.org/#2} {doi:#2}\else \href {http://dx.doi.org/#2} {#1}\fi
  \endgroup}
\def\mn@eprint#1#2{\mn@eprint@#1:#2::\@nil}
\def\mn@eprint@arXiv#1{\href {http://arxiv.org/abs/#1} {{\tt arXiv:#1}}}
\def\mn@eprint@dblp#1{\href {http://dblp.uni-trier.de/rec/bibtex/#1.xml}
  {dblp:#1}}
\def\mn@eprint@#1:#2:#3:#4\@nil{\def\@tempa {#1}\def\@tempb {#2}\def\@tempc
  {#3}\ifx \@tempc \@empty \let \@tempc \@tempb \let \@tempb \@tempa \fi \ifx
  \@tempb \@empty \def\@tempb {arXiv}\fi \@ifundefined
  {mn@eprint@\@tempb}{\@tempb:\@tempc}{\expandafter \expandafter \csname
  mn@eprint@\@tempb\endcsname \expandafter{\@tempc}}}

\bibitem[\protect\citeauthoryear{{Abramowicz} \& {Klu{\'z}niak}}{{Abramowicz}
  \& {Klu{\'z}niak}}{2001}]{Abramowicz01}
{Abramowicz} M.~A.,  {Klu{\'z}niak} W.,  2001, \mn@doi [\aap]
  {10.1051/0004-6361:20010791}, \href
  {http://adsabs.harvard.edu/abs/2001A%26A...374L..19A} {374, L19}

\bibitem[\protect\citeauthoryear{{Abramowicz}, {Jaroszynski}  \&
  {Sikora}}{{Abramowicz} et~al.}{1978}]{Abramowicz1978}
{Abramowicz} M.,  {Jaroszynski} M.,   {Sikora} M.,  1978, \aap, \href
  {http://adsabs.harvard.edu/abs/1978A%26A....63..221A} {63, 221}

\bibitem[\protect\citeauthoryear{{Abramowicz}, {Czerny}, {Lasota}  \&
  {Szuszkiewicz}}{{Abramowicz} et~al.}{1988}]{Abramowicz1988}
{Abramowicz} M.~A.,  {Czerny} B.,  {Lasota} J.~P.,   {Szuszkiewicz} E.,  1988,
  \mn@doi [\apj] {10.1086/166683}, \href
  {http://adsabs.harvard.edu/abs/1988ApJ...332..646A} {332, 646}

\bibitem[\protect\citeauthoryear{{Altamirano} \& {Belloni}}{{Altamirano} \&
  {Belloni}}{2012}]{Altamirano2012}
{Altamirano} D.,  {Belloni} T.,  2012, \mn@doi [\apjl]
  {10.1088/2041-8205/747/1/L4}, \href
  {http://adsabs.harvard.edu/abs/2012ApJ...747L...4A} {747, L4}

\bibitem[\protect\citeauthoryear{{Anninos}, {Fragile}  \&
  {Salmonson}}{{Anninos} et~al.}{2005}]{Anninos05}
{Anninos} P.,  {Fragile} P.~C.,   {Salmonson} J.~D.,  2005, \mn@doi [\apj]
  {10.1086/497294}, \href {http://adsabs.harvard.edu/abs/2005ApJ...635..723A}
  {635, 723}

\bibitem[\protect\citeauthoryear{{Belloni}, {Klein-Wolt}, {M{\'e}ndez}, {van
  der Klis}  \& {van Paradijs}}{{Belloni} et~al.}{2000}]{Belloni2000}
{Belloni} T.,  {Klein-Wolt} M.,  {M{\'e}ndez} M.,  {van der Klis} M.,   {van
  Paradijs} J.,  2000, \aap, \href
  {http://adsabs.harvard.edu/abs/2000A%26A...355..271B} {355, 271}

\bibitem[\protect\citeauthoryear{{Belloni}, {M{\'e}ndez}  \&
  {S{\'a}nchez-Fern{\'a}ndez}}{{Belloni} et~al.}{2001}]{Belloni2001}
{Belloni} T.,  {M{\'e}ndez} M.,   {S{\'a}nchez-Fern{\'a}ndez} C.,  2001,
  \mn@doi [\aap] {10.1051/0004-6361:20010480}, \href
  {http://adsabs.harvard.edu/abs/2001A%26A...372..551B} {372, 551}

\bibitem[\protect\citeauthoryear{{Belloni}, {Sanna}  \& {M{\'e}ndez}}{{Belloni}
  et~al.}{2012}]{Belloni2012}
{Belloni} T.~M.,  {Sanna} A.,   {M{\'e}ndez} M.,  2012, \mn@doi [\mnras]
  {10.1111/j.1365-2966.2012.21634.x}, \href
  {http://adsabs.harvard.edu/abs/2012MNRAS.426.1701B} {426, 1701}

\bibitem[\protect\citeauthoryear{{Blaes}, {Arras}  \& {Fragile}}{{Blaes}
  et~al.}{2006}]{Blaes06}
{Blaes} O.~M.,  {Arras} P.,   {Fragile} P.~C.,  2006, \mn@doi [\mnras]
  {10.1111/j.1365-2966.2006.10370.x}, \href
  {http://adsabs.harvard.edu/abs/2006MNRAS.369.1235B} {369, 1235}

\bibitem[\protect\citeauthoryear{{Blaes}, {Krolik}, {Hirose}  \&
  {Shabaltas}}{{Blaes} et~al.}{2011}]{Blaes2011}
{Blaes} O.,  {Krolik} J.~H.,  {Hirose} S.,   {Shabaltas} N.,  2011, \mn@doi
  [\apj] {10.1088/0004-637X/733/2/110}, \href
  {http://adsabs.harvard.edu/abs/2011ApJ...733..110B} {733, 110}

\bibitem[\protect\citeauthoryear{{Bollimpalli} \& {Klu{\'z}niak}}{{Bollimpalli}
  \& {Klu{\'z}niak}}{2017}]{deepika2017}
{Bollimpalli} D.~A.,  {Klu{\'z}niak} W.,  2017, \mn@doi [\mnras]
  {10.1093/mnras/stx2140}, \href
  {http://adsabs.harvard.edu/abs/2017MNRAS.472.3298B} {472, 3298}

\bibitem[\protect\citeauthoryear{{Bursa}, {Abramowicz}, {Karas}  \&
  {Klu{\'z}niak}}{{Bursa} et~al.}{2004}]{Bursa04}
{Bursa} M.,  {Abramowicz} M.~A.,  {Karas} V.,   {Klu{\'z}niak} W.,  2004,
  \mn@doi [\apjl] {10.1086/427167}, \href
  {http://adsabs.harvard.edu/abs/2004ApJ...617L..45B} {617, L45}

\bibitem[\protect\citeauthoryear{{Fragile}, {Gillespie}, {Monahan}, {Rodriguez}
   \& {Anninos}}{{Fragile} et~al.}{2012}]{Fragile12}
{Fragile} P.~C.,  {Gillespie} A.,  {Monahan} T.,  {Rodriguez} M.,   {Anninos}
  P.,  2012, \mn@doi [\apjs] {10.1088/0067-0049/201/2/9}, \href
  {http://adsabs.harvard.edu/abs/2012ApJS..201....9F} {201, 9}

\bibitem[\protect\citeauthoryear{{Fragile}, {Olejar}  \& {Anninos}}{{Fragile}
  et~al.}{2014}]{Fragile14}
{Fragile} P.~C.,  {Olejar} A.,   {Anninos} P.,  2014, \mn@doi [\apj]
  {10.1088/0004-637X/796/1/22}, \href
  {http://adsabs.harvard.edu/abs/2014ApJ...796...22F} {796, 22}

\bibitem[\protect\citeauthoryear{{Fragile}, {Straub}  \& {Blaes}}{{Fragile}
  et~al.}{2016}]{Fragile16}
{Fragile} P.~C.,  {Straub} O.,   {Blaes} O.,  2016, \mn@doi [\mnras]
  {10.1093/mnras/stw1428}, \href
  {http://adsabs.harvard.edu/abs/2016MNRAS.461.1356F} {461, 1356}

\bibitem[\protect\citeauthoryear{{Fragile}, {Etheridge}, {Anninos}, {Mishra}
  \& {Kluzniak}}{{Fragile} et~al.}{2018}]{Fragile2018}
{Fragile} P.~C.,  {Etheridge} S.~M.,  {Anninos} P.,  {Mishra} B.,   {Kluzniak}
  W.,  2018, preprint, \href
  {http://adsabs.harvard.edu/abs/2018arXiv180306423F} {} (\mn@eprint {arXiv}
  {1803.06423})

\bibitem[\protect\citeauthoryear{{Hirose}, {Krolik}  \& {Blaes}}{{Hirose}
  et~al.}{2009}]{Hirose09}
{Hirose} S.,  {Krolik} J.~H.,   {Blaes} O.,  2009, \mn@doi [\apj]
  {10.1088/0004-637X/691/1/16}, \href
  {http://adsabs.harvard.edu/abs/2009ApJ...691...16H} {691, 16}

\bibitem[\protect\citeauthoryear{{Hogg} \& {Reynolds}}{{Hogg} \&
  {Reynolds}}{2018}]{Hogg2018}
{Hogg} J.~D.,  {Reynolds} C.~S.,  2018, preprint, \href
  {http://adsabs.harvard.edu/abs/2018arXiv180105836H} {} (\mn@eprint {arXiv}
  {1801.05836})

\bibitem[\protect\citeauthoryear{{Ingram}, {Done}  \& {Fragile}}{{Ingram}
  et~al.}{2009}]{Ingram2009}
{Ingram} A.,  {Done} C.,   {Fragile} P.~C.,  2009, \mn@doi [\mnras]
  {10.1111/j.1745-3933.2009.00693.x}, \href
  {http://adsabs.harvard.edu/abs/2009MNRAS.397L.101I} {397, L101}

\bibitem[\protect\citeauthoryear{{Jensen}, {Christensen}  \&
  {Fogedby}}{{Jensen} et~al.}{1989}]{Jensen1989}
{Jensen} H.~J.,  {Christensen} K.,   {Fogedby} H.~C.,  1989, \mn@doi [\prb]
  {10.1103/PhysRevB.40.7425}, \href
  {http://adsabs.harvard.edu/abs/1989PhRvB..40.7425J} {40, 7425}

\bibitem[\protect\citeauthoryear{{Jiang}, {Stone}  \& {Davis}}{{Jiang}
  et~al.}{2013}]{Jiang13}
{Jiang} Y.-F.,  {Stone} J.~M.,   {Davis} S.~W.,  2013, \mn@doi [\apj]
  {10.1088/0004-637X/778/1/65}, \href
  {http://adsabs.harvard.edu/abs/2013ApJ...778...65J} {778, 65}

\bibitem[\protect\citeauthoryear{{Kato}}{{Kato}}{2001}]{Kato2001}
{Kato} S.,  2001, \mn@doi [\pasj] {10.1093/pasj/53.1.1}, \href
  {http://adsabs.harvard.edu/abs/2001PASJ...53....1K} {53, 1}

\bibitem[\protect\citeauthoryear{Kato}{Kato}{2016}]{katoBook}
Kato S.,  2016, Oscillations of Disks.
~0067 Vol. 437, Springer Japan

\bibitem[\protect\citeauthoryear{{Klu{\'z}niak}}{{Klu{\'z}niak}}{1998}]{Kluzniak1998}
{Klu{\'z}niak} W.,  1998, \mn@doi [\apjl] {10.1086/311748}, \href
  {http://adsabs.harvard.edu/abs/1998ApJ...509L..37K} {509, L37}

\bibitem[\protect\citeauthoryear{{Klu{\'z}niak}}{{Klu{\'z}niak}}{2005}]{Kluzniak05}
{Klu{\'z}niak} W.,  2005, \mn@doi [Astronomische Nachrichten]
  {10.1002/asna.200510420}, \href
  {http://adsabs.harvard.edu/abs/2005AN....326..820K} {326, 820}

\bibitem[\protect\citeauthoryear{{Klu\'zniak} \& {Abramowicz}}{{Klu\'zniak} \&
  {Abramowicz}}{2001}]{Kluzniak01}
{Klu\'zniak} W.,  {Abramowicz} M.~A.,  2001, Acta Physica Polonica B, \href
  {http://adsabs.harvard.edu/abs/2001AcPPB..32.3605K} {32, 3605}

\bibitem[\protect\citeauthoryear{{Klu\'zniak} \& {Abramowicz}}{{Klu\'zniak} \&
  {Abramowicz}}{2002}]{Kluzniak02}
{Klu\'zniak} W.,  {Abramowicz} M.~A.,  2002, ArXiv Astrophysics e-prints, \href
  {http://adsabs.harvard.edu/abs/2002astro.ph..3314K} {}

\bibitem[\protect\citeauthoryear{{Klu\'zniak} \& {Kita}}{{Klu\'zniak} \&
  {Kita}}{2000}]{KK00}
{Klu\'zniak} W.,  {Kita} D.,  2000, ArXiv Astrophysics e-prints, \href
  {http://adsabs.harvard.edu/abs/2000astro.ph..6266K} {}

\bibitem[\protect\citeauthoryear{{Klu\'zniak}, {Michelson}  \&
  {Wagoner}}{{Klu\'zniak} et~al.}{1990}]{Kluzniak90}
{Klu\'zniak} W.,  {Michelson} P.,   {Wagoner} R.~V.,  1990, \mn@doi [\apj]
  {10.1086/169006}, \href {http://adsabs.harvard.edu/abs/1990ApJ...358..538K}
  {358, 538}

\bibitem[\protect\citeauthoryear{{Lightman} \& {Eardley}}{{Lightman} \&
  {Eardley}}{1974}]{Lightman74}
{Lightman} A.~P.,  {Eardley} D.~M.,  1974, \mn@doi [\apjl] {10.1086/181377},
  \href {http://adsabs.harvard.edu/abs/1974ApJ...187L...1L} {187, L1}

\bibitem[\protect\citeauthoryear{{Mazur}, {Zanotti}, {S{\c a}dowski}, {Mishra}
  \& {Klu{\'z}niak}}{{Mazur} et~al.}{2016}]{Mazur16}
{Mazur} G.~P.,  {Zanotti} O.,  {S{\c a}dowski} A.,  {Mishra} B.,
  {Klu{\'z}niak} W.,  2016, \mn@doi [\mnras] {10.1093/mnras/stv2890}, \href
  {http://adsabs.harvard.edu/abs/2016MNRAS.456.3245M} {456, 3245}

\bibitem[\protect\citeauthoryear{{Meekins}, {Wood}, {Hedler}, {Byram},
  {Yentis}, {Chubb}  \& {Friedman}}{{Meekins} et~al.}{1984}]{Meekins84}
{Meekins} J.~F.,  {Wood} K.~S.,  {Hedler} R.~L.,  {Byram} E.~T.,  {Yentis}
  D.~J.,  {Chubb} T.~A.,   {Friedman} H.,  1984, \mn@doi [\apj]
  {10.1086/161793}, \href {http://adsabs.harvard.edu/abs/1984ApJ...278..288M}
  {278, 288}

\bibitem[\protect\citeauthoryear{{Mihalas} \& {Mihalas}}{{Mihalas} \&
  {Mihalas}}{1984}]{Mihalas84}
{Mihalas} D.,  {Mihalas} B.~W.,  1984, {Foundations of radiation hydrodynamics}

\bibitem[\protect\citeauthoryear{{Miller} et~al.,}{{Miller}
  et~al.}{2001}]{Miller2001}
{Miller} J.~M.,  et~al., 2001, \mn@doi [\apj] {10.1086/324027}, \href
  {http://adsabs.harvard.edu/abs/2001ApJ...563..928M} {563, 928}

\bibitem[\protect\citeauthoryear{{Mishra}, {Fragile}, {Johnson}  \&
  {Klu{\'z}niak}}{{Mishra} et~al.}{2016}]{Mishra16}
{Mishra} B.,  {Fragile} P.~C.,  {Johnson} L.~C.,   {Klu{\'z}niak} W.,  2016,
  \mn@doi [\mnras] {10.1093/mnras/stw2245}, \href
  {http://adsabs.harvard.edu/abs/2016MNRAS.463.3437M} {463, 3437}

\bibitem[\protect\citeauthoryear{{Mishra}, {Vincent}, {Manousakis}, {Fragile},
  {Paumard}  \& {Klu{\'z}niak}}{{Mishra} et~al.}{2017}]{Mishra17}
{Mishra} B.,  {Vincent} F.~H.,  {Manousakis} A.,  {Fragile} P.~C.,  {Paumard}
  T.,   {Klu{\'z}niak} W.,  2017, \mn@doi [\mnras] {10.1093/mnras/stx299},
  \href {http://adsabs.harvard.edu/abs/2017MNRAS.467.4036M} {467, 4036}

\bibitem[\protect\citeauthoryear{{Morales Teixeira}, {Avara}  \&
  {McKinney}}{{Morales Teixeira} et~al.}{2017}]{Morales2017}
{Morales Teixeira} D.,  {Avara} M.~J.,   {McKinney} J.~C.,  2017, preprint,
  \href {http://adsabs.harvard.edu/abs/2017arXiv170608982M} {} (\mn@eprint
  {arXiv} {1706.08982})

\bibitem[\protect\citeauthoryear{{Morgan}, {Remillard}  \& {Greiner}}{{Morgan}
  et~al.}{1997}]{Morgan97}
{Morgan} E.~H.,  {Remillard} R.~A.,   {Greiner} J.,  1997, \mn@doi [\apj]
  {10.1086/304191}, \href {http://adsabs.harvard.edu/abs/1997ApJ...482..993M}
  {482, 993}

\bibitem[\protect\citeauthoryear{{Novikov} \& {Thorne}}{{Novikov} \&
  {Thorne}}{1973}]{Novikov73}
{Novikov} I.~D.,  {Thorne} K.~S.,  1973, in {Dewitt} C.,  {Dewitt} B.~S.,  eds,
  Black Holes (Les Astres Occlus). pp 343--450

\bibitem[\protect\citeauthoryear{{O'Neill}, {Reynolds}  \& {Miller}}{{O'Neill}
  et~al.}{2009}]{ONeill09}
{O'Neill} S.~M.,  {Reynolds} C.~S.,   {Miller} M.~C.,  2009, \mn@doi [\apj]
  {10.1088/0004-637X/693/2/1100}, \href
  {http://adsabs.harvard.edu/abs/2009ApJ...693.1100O} {693, 1100}

\bibitem[\protect\citeauthoryear{{Piran}}{{Piran}}{1978}]{Piran78}
{Piran} T.,  1978, \mn@doi [\apj] {10.1086/156069}, \href
  {http://adsabs.harvard.edu/abs/1978ApJ...221..652P} {221, 652}

\bibitem[\protect\citeauthoryear{{Pringle} \& {Rees}}{{Pringle} \&
  {Rees}}{1972}]{Pringle1972}
{Pringle} J.~E.,  {Rees} M.~J.,  1972, \aap, \href
  {http://adsabs.harvard.edu/abs/1972A%26A....21....1P} {21, 1}

\bibitem[\protect\citeauthoryear{{Remillard} \& {McClintock}}{{Remillard} \&
  {McClintock}}{2006}]{Remillard06}
{Remillard} R.~A.,  {McClintock} J.~E.,  2006, \mn@doi [\araa]
  {10.1146/annurev.astro.44.051905.092532}, \href
  {http://adsabs.harvard.edu/abs/2006ARA%26A..44...49R} {44, 49}

\bibitem[\protect\citeauthoryear{{Remillard}, {Morgan}, {McClintock}, {Bailyn}
  \& {Orosz}}{{Remillard} et~al.}{1999}]{Remillard99}
{Remillard} R.~A.,  {Morgan} E.~H.,  {McClintock} J.~E.,  {Bailyn} C.~D.,
  {Orosz} J.~A.,  1999, \mn@doi [\apj] {10.1086/307606}, \href
  {http://adsabs.harvard.edu/abs/1999ApJ...522..397R} {522, 397}

\bibitem[\protect\citeauthoryear{{Remillard}, {Muno}, {McClintock}  \&
  {Orosz}}{{Remillard} et~al.}{2002}]{Remillard02}
{Remillard} R.~A.,  {Muno} M.~P.,  {McClintock} J.~E.,   {Orosz} J.~A.,  2002,
  \mn@doi [\apj] {10.1086/343791}, \href
  {http://adsabs.harvard.edu/abs/2002ApJ...580.1030R} {580, 1030}

\bibitem[\protect\citeauthoryear{{Reynolds} \& {Miller}}{{Reynolds} \&
  {Miller}}{2009}]{Reynolds09}
{Reynolds} C.~S.,  {Miller} M.~C.,  2009, \mn@doi [\apj]
  {10.1088/0004-637X/692/1/869}, \href
  {http://adsabs.harvard.edu/abs/2009ApJ...692..869R} {692, 869}

\bibitem[\protect\citeauthoryear{{Rezzolla}, {Yoshida}, {Maccarone}  \&
  {Zanotti}}{{Rezzolla} et~al.}{2003}]{Rezzolla03}
{Rezzolla} L.,  {Yoshida} S.,  {Maccarone} T.~J.,   {Zanotti} O.,  2003,
  \mn@doi [\mnras] {10.1046/j.1365-8711.2003.07018.x}, \href
  {http://adsabs.harvard.edu/abs/2003MNRAS.344L..37R} {344, L37}

\bibitem[\protect\citeauthoryear{{S{\c a}dowski}}{{S{\c
  a}dowski}}{2016}]{Sadowski16}
{S{\c a}dowski} A.,  2016, \mn@doi [\mnras] {10.1093/mnras/stw913}, \href
  {http://adsabs.harvard.edu/abs/2016MNRAS.459.4397S} {459, 4397}

\bibitem[\protect\citeauthoryear{{S{\c a}dowski}, {Narayan}, {Tchekhovskoy}  \&
  {Zhu}}{{S{\c a}dowski} et~al.}{2013}]{Sadowski13}
{S{\c a}dowski} A.,  {Narayan} R.,  {Tchekhovskoy} A.,   {Zhu} Y.,  2013,
  \mn@doi [\mnras] {10.1093/mnras/sts632}, \href
  {http://adsabs.harvard.edu/abs/2013MNRAS.429.3533S} {429, 3533}

\bibitem[\protect\citeauthoryear{{Shakura} \& {Sunyaev}}{{Shakura} \&
  {Sunyaev}}{1973}]{Shakura73}
{Shakura} N.~I.,  {Sunyaev} R.~A.,  1973, \aap, \href
  {http://adsabs.harvard.edu/abs/1973A%26A....24..337S} {24, 337}

\bibitem[\protect\citeauthoryear{{Shakura} \& {Sunyaev}}{{Shakura} \&
  {Sunyaev}}{1976}]{Shakura76}
{Shakura} N.~I.,  {Sunyaev} R.~A.,  1976, \mn@doi [\mnras]
  {10.1093/mnras/175.3.613}, \href
  {http://adsabs.harvard.edu/abs/1976MNRAS.175..613S} {175, 613}

\bibitem[\protect\citeauthoryear{{Silbergleit}, {Wagoner}  \&
  {Ortega-Rodr{\'{\i}}guez}}{{Silbergleit} et~al.}{2001}]{silb2001}
{Silbergleit} A.~S.,  {Wagoner} R.~V.,   {Ortega-Rodr{\'{\i}}guez} M.,  2001,
  \mn@doi [\apj] {10.1086/318659}, \href
  {http://adsabs.harvard.edu/abs/2001ApJ...548..335S} {548, 335}

\bibitem[\protect\citeauthoryear{{Stella} \& {Vietri}}{{Stella} \&
  {Vietri}}{1998}]{Stella98}
{Stella} L.,  {Vietri} M.,  1998, \mn@doi [\apjl] {10.1086/311075}, \href
  {http://adsabs.harvard.edu/abs/1998ApJ...492L..59S} {492, L59}

\bibitem[\protect\citeauthoryear{{Stella} \& {Vietri}}{{Stella} \&
  {Vietri}}{1999}]{Stella99}
{Stella} L.,  {Vietri} M.,  1999, \mn@doi [Physical Review Letters]
  {10.1103/PhysRevLett.82.17}, \href
  {http://adsabs.harvard.edu/abs/1999PhRvL..82...17S} {82, 17}

\bibitem[\protect\citeauthoryear{{Strohmayer}}{{Strohmayer}}{2001}]{Strohmayer01}
{Strohmayer} T.~E.,  2001, \mn@doi [\apjl] {10.1086/320258}, \href
  {http://adsabs.harvard.edu/abs/2001ApJ...552L..49S} {552, L49}

\bibitem[\protect\citeauthoryear{{Wagoner}}{{Wagoner}}{1999}]{Wagoner99}
{Wagoner} R.~V.,  1999, \mn@doi [\physrep] {10.1016/S0370-1573(98)00104-5},
  \href {http://adsabs.harvard.edu/abs/1999PhR...311..259W} {311, 259}

\bibitem[\protect\citeauthoryear{{Zanotti}, {Rezzolla}  \& {Font}}{{Zanotti}
  et~al.}{2003}]{Zanotti03}
{Zanotti} O.,  {Rezzolla} L.,   {Font} J.~A.,  2003, \mn@doi [\mnras]
  {10.1046/j.1365-8711.2003.06474.x}, \href
  {http://adsabs.harvard.edu/abs/2003MNRAS.341..832Z} {341, 832}

\bibitem[\protect\citeauthoryear{{Zanotti}, {Font}, {Rezzolla}  \&
  {Montero}}{{Zanotti} et~al.}{2005}]{Zanotti05}
{Zanotti} O.,  {Font} J.~A.,  {Rezzolla} L.,   {Montero} P.~J.,  2005, \mn@doi
  [\mnras] {10.1111/j.1365-2966.2004.08567.x}, \href
  {http://adsabs.harvard.edu/abs/2005MNRAS.356.1371Z} {356, 1371}

\bibitem[\protect\citeauthoryear{{van der Klis}}{{van der
  Klis}}{2000}]{vanderklis2000}
{van der Klis} M.,  2000, \mn@doi [\araa] {10.1146/annurev.astro.38.1.717},
  \href {http://adsabs.harvard.edu/abs/2000ARA%26A..38..717V} {38, 717}

\makeatother
\end{thebibliography}

%%%%%%%%%%%%%%%%%%%%%%%%%%%%%%%%%%%%%%%%%%%%%%%%%%

%%%%%%%%%%%%%%%%% APPENDICES %%%%%%%%%%%%%%%%%%%%%

\end{document}